\documentstyle[12pt]{article}
\advance \topmargin by -\headheight
\advance \topmargin by -\headsep

\evensidemargin \oddsidemargin
\begin{document}

\topmargin -.6in


\def\rf#1{(\ref{eq:#1})}
\def\lab#1{\label{eq:#1}}
\def\nonu{\nonumber}
\def\br{\begin{eqnarray}}
\def\er{\end{eqnarray}}
\def\be{\begin{equation}}
\def\ee{\end{equation}}
\def\eq{\!\!\!\! &=& \!\!\!\! }
\def\foot#1{\footnotemark\footnotetext{#1}}
\def\lb{\lbrack}
\def\rb{\rbrack}
\def\llangle{\left\langle}
\def\rrangle{\right\rangle}
\def\blangle{\Bigl\langle}
\def\brangle{\Bigr\rangle}
\def\llb{\left\lbrack}
\def\rrb{\right\rbrack}
\def\Blb{\Bigl\lbrack}
\def\Brb{\Bigr\rbrack}
\def\lcurl{\left\{}
\def\rcurl{\right\}}
\def\({\left(}
\def\){\right)}
\def\v{\vert}                     
\def\bv{\bigm\vert}               
\def\Bgv{\;\Bigg\vert}            
\def\bgv{\bigg\vert}              
\def\lskip{\vskip\baselineskip\vskip-\parskip\noindent}
\def\mskp{\par\vskip 0.3cm \par\noindent}
\def\sskp{\par\vskip 0.15cm \par\noindent}
\def\bc{\begin{center}}
\def\ec{\end{center}}
\def\Lbf#1{{\Large {\bf {#1}}}}
\def\lbf#1{{\large {\bf {#1}}}}
\relax


\def\tr{\mathop{\rm tr}}                  
\def\Tr{\mathop{\rm Tr}}                  
\newcommand\partder[2]{{{\partial {#1}}\over{\partial {#2}}}}
\newcommand\funcder[2]{{{\delta {#1}}\over{\delta {#2}}}}
\newcommand\Bil[2]{\Bigl\langle {#1} \Bigg\vert {#2} \Bigr\rangle}  
\newcommand\bil[2]{\left\langle {#1} \bigg\vert {#2} \right\rangle} 
\newcommand\me[2]{\left\langle {#1}\right|\left. {#2} \right\rangle} 

\newcommand\sbr[2]{\left\lbrack\,{#1}\, ,\,{#2}\,\right\rbrack} 
\newcommand\Sbr[2]{\Bigl\lbrack\,{#1}\, ,\,{#2}\,\Bigr\rbrack} 
\newcommand\pbr[2]{\{\,{#1}\, ,\,{#2}\,\}}       
\newcommand\Pbr[2]{\Bigl\{ \,{#1}\, ,\,{#2}\,\Bigr\}}  
\newcommand\pbbr[2]{\lcurl\,{#1}\, ,\,{#2}\,\rcurl}  


\def\a{\alpha}
\def\b{\beta}
\def\c{\chi}
\def\d{\delta}
\def\D{\Delta}
\def\eps{\epsilon}
\def\vareps{\varepsilon}
\def\g{\gamma}
\def\G{\Gamma}
\def\grad{\nabla}
\def\h{{1\over 2}}
\def\l{\lambda}
\def\L{\Lambda}
\def\m{\mu}
\def\n{\nu}
\def\ov{\over}
\def\om{\omega}
\def\O{\Omega}
\def\p{\phi}
\def\P{\Phi}
\def\pa{\partial}
\def\pr{\prime}
\def\ra{\rightarrow}
\def\s{\sigma}
\def\S{\Sigma}
\def\t{\tau}
\def\th{\theta}
\def\Th{\Theta}
\def\z{\zeta}
\def\ti{\tilde}
\def\wti{\widetilde}
\newcommand\sumi[1]{\sum_{#1}^{\infty}}   


\def\cA{{\cal A}}
\def\cB{{\cal B}}
\def\cC{{\cal C}}
\def\cD{{\cal D}}
\def\cE{{\cal E}}
\def\cL{{\cal L}}
\def\cM{{\cal M}}
\def\cN{{\cal N}}
\def\cP{{\cal P}}
\def\cQ{{\cal Q}}
\def\cR{{\cal R}}
\def\cS{{\cal S}}
\def\cU{{\cal U}}
\def\cV{{\cal V}}
\def\cW{{\cal W}}
\def\cY{{\cal Y}}


\def\phanta{\phantom{aaaaaaaaaaaaaaa}}
\def\phantb{\phantom{aaaaaaaaaaaaaaaaaaaaaaaaa}}
\def\phantc{\phantom{aaaaaaaaaaaaaaaaaaaaaaaaaaaaaaaaaaa}}


\def\rlx{\relax\leavevmode}
\def\inbar{\vrule height1.5ex width.4pt depth0pt}
\def\IZ{\rlx\hbox{\sf Z\kern-.4em Z}}
\def\IR{\rlx\hbox{\rm I\kern-.18em R}}


\def\mark{\noindent{\bf Remark.}\quad}
\def\prop{\noindent{\bf Proposition.}\quad}
\def\theor{\noindent{\bf Theorem.}\quad}
\def\name{\noindent{\bf Definition.}\quad}
\def\exam{\noindent{\bf Example.}\quad}
\def\proof{\noindent{\bf Proof.}\quad}


\def\Winf{{\bf W_\infty}}               
\def\Win1{{\bf W_{1+\infty}}}           
\def\Winft#1{{\bf W_\infty^{\geq {#1}}}}    
\def\winf{{\bf w_\infty}}
\def\win1{{\bf w_{1+\infty}}}
\def\hWinf{{\bf {\hat W}_{\infty}}}        

\def\sj{{\jmath}{}}
\def\bsj{{\bar \jmath}{}}
\def\bp{{\bar \p}}
\def\faa{Fa\'a di Bruno~}

\def\AM#1{A^{(M)}_{#1}}
\def\BM#1{B^{(M)}_{#1}}
\def\Xb{X(b_{M})}
\def\Yb{Y(b_{M})}
\def\Xbo{X_{(0)}(b_{M})}
\def\Ybo{Y_{(0)}(b_{M})}
\newcommand{\nit}{\noindent}
\newcommand{\ct}[1]{\cite{#1}}
\newcommand{\bi}[1]{\bibitem{#1}}
%
%
%
\newcommand\PRL[3]{{\sl Phys. Rev. Lett.} {\bf#1} (#2) #3}
\newcommand\NPB[3]{{\sl Nucl. Phys.} {\bf B#1} (#2) #3}
\newcommand\NPBFS[4]{{\sl Nucl. Phys.} {\bf B#2} [FS#1] (#3) #4}
\newcommand\CMP[3]{{\sl Commun. Math. Phys.} {\bf #1} (#2) #3}
\newcommand\PRD[3]{{\sl Phys. Rev.} {\bf D#1} (#2) #3}
\newcommand\PLA[3]{{\sl Phys. Lett.} {\bf #1A} (#2) #3}
\newcommand\PLB[3]{{\sl Phys. Lett.} {\bf #1B} (#2) #3}
\newcommand\JMP[3]{{\sl J. Math. Phys.} {\bf #1} (#2) #3}
\newcommand\PTP[3]{{\sl Prog. Theor. Phys.} {\bf #1} (#2) #3}
\newcommand\SPTP[3]{{\sl Suppl. Prog. Theor. Phys.} {\bf #1} (#2) #3}
\newcommand\AoP[3]{{\sl Ann. of Phys.} {\bf #1} (#2) #3}
\newcommand\RMP[3]{{\sl Rev. Mod. Phys.} {\bf #1} (#2) #3}
\newcommand\PR[3]{{\sl Phys. Reports} {\bf #1} (#2) #3}
\newcommand\FAP[3]{{\sl Funkt. Anal. Prilozheniya} {\bf #1} (#2) #3}
\newcommand\FAaIA[3]{{\sl Functional Analysis and Its Application} {\bf #1}
(#2) #3}
\newcommand\LMP[3]{{\sl Letters in Math. Phys.} {\bf #1} (#2) #3}
\newcommand\IJMPA[3]{{\sl Int. J. Mod. Phys.} {\bf A#1} (#2) #3}
\newcommand\TMP[3]{{\sl Theor. Mat. Phys.} {\bf #1} (#2) #3}
\newcommand\JPA[3]{{\sl J. Physics} {\bf A#1} (#2) #3}
\newcommand\JSM[3]{{\sl J. Soviet Math.} {\bf #1} (#2) #3}
\newcommand\MPLA[3]{{\sl Mod. Phys. Lett.} {\bf A#1} (#2) #3}
\newcommand\JETP[3]{{\sl Sov. Phys. JETP} {\bf #1} (#2) #3}
\newcommand\JETPL[3]{{\sl  Sov. Phys. JETP Lett.} {\bf #1} (#2) #3}
\newcommand\PHSA[3]{{\sl Physica} {\bf A#1} (#2) #3}
\newcommand\PHSD[3]{{\sl Physica} {\bf D#1} (#2) #3}
\begin{titlepage}
\vspace*{-1cm}
\noindent
January, 1994 \hfill{BGU-94 / 1 / January- PH}\\
\phantom{bla}
\hfill{Revised April 1994}\\
\phantom{bla}
\hfill{UICHEP-TH/94-2}\\
\phantom{bla}
\hfill{hep-th/9401058}
\\
\begin{center}
{\large\bf Hamiltonian Structures of the Multi-Boson KP Hierarchies, \\
Abelianization and Lattice Formulation}
\end{center}
\vskip .3in
\begin{center}
{ H. Aratyn\footnotemark
\footnotetext{Work supported in part by the U.S. Department of Energy
under contract DE-FG02-84ER40173}}
\par \vskip .1in \noindent
Department of Physics \\
University of Illinois at Chicago\\
845 W. Taylor St.\\
Chicago, IL 60607-7059, {\em e-mail}:
u23325@uicvm \\
\par \vskip .3in
{ E. Nissimov$^{\,2}$  and S. Pacheva \footnotemark
\footnotetext{On leave from: Institute of Nuclear Research and Nuclear
Energy, Boul. Tsarigradsko Chausee 72, BG-1784 $\;$Sofia,
Bulgaria. }}
\par \vskip .1in \noindent
Department of Physics, Ben-Gurion University of the Negev \\
Box 653, IL-84105 $\;$Beer Sheva, Israel \\
{\em e-mail}: emil@bguvms, svetlana@bguvms
\par \vskip .3in
\end{center}

\begin{abstract}

We present a new  form of the multi-boson reduction of KP hierarchy
with Lax operator written in terms of boson fields abelianizing the
second Hamiltonian structure.
This extends the classical Miura transformation and the
Kupershmidt-Wilson theorem from the (m)KdV to the KP case.
A remarkable relationship is uncovered between the higher Hamiltonian
structures and the corresponding Miura transformations of KP hierarchy,
on one hand, and the discrete integrable models living on {\em refinements}
of the original lattice connected with the underlying multi-matrix models,
on the other hand.
For the second KP Hamiltonian structure, worked out in details, this amounts to
finding a series of representations of the nonlinear $\hWinf$ algebra
in terms of arbitrary finite number of canonical pairs of free fields.

\end{abstract}

\end{titlepage}

\noindent
{\large {\bf 1. Introduction}}
\lskip
Multi-boson Kadomtsev-Petviashvili (KP) hierarchies are integrable systems
of a very unique structure.
Since their appearance as generalizations of two- and four-boson KP hierarchies
\ct{R82,LM79,BAK85,YW9111,2boson,ANPV} in the context of the matrix models of
strings \ct{BX9204,BX9209,BX9212,BX9305} several of their intriguing
properties have been revealed and studied.

Multi-boson KP hierarchies are consistent {\em Poisson reductions} of the
standard full (infini\-te\-ly-many-field)
KP hierarchy within the $R$-matrix scheme \ct{ANP93}. Let us note, that
within the Lax formulation of integrable Hamiltonian systems, restrictions
of the pertinent
Lax operator to a submanifold {\em do not} necessarily lead to consistent
restrictions of the corresponding Poisson structures \ct{Morosi}. Thus, the
proof of consistency of the Poisson reductions is a necessary step
in the construction of multi-boson KP hierarchies. In ref.\ct{ANP93}
the first KP Hamiltonian structure was formulated in
terms of Darboux-Poisson  coordinate pairs, {\sl i.e.}, canonical pairs of
free fields. This resulted in a series of representations of $\Win1\,$
algebra made out of arbitrary even number of free boson fields.

In this letter we first present a new formulation of the multi-boson
reduction of KP hierarchy in terms of a different set of boson fields
abelianizing the second Hamiltonian structure.
This extends the classical Miura transformation and the
Kupershmidt-Wilson theorem \cite{Dickey} from the mKdV
(mofidied Korteveg-de Vries) to the KP case.
Within the KP context, it is a generalization of the Miura
mapping between linear and quadratic two-boson KP hierarchies \ct{AFGMZ} to
arbitrary multi-boson KP hierarchies.

One of the unique features of multi-boson KP hierarchies is their connection
to the Toda lattices originating from the multi-matrix models
\ct{BX9204,BX9209}, and the corresponding invariance
under discrete symmetries \ct{LSY93,discrete,similar}.
The discrete symmetries, being implemented by a similarity transformation of
the Lax operator, are canonical transformations leaving all the Hamiltonian
structures form-invariant \ct{similar}.

In this letter we next exhibit
a remarkable relationship  between the higher Hamiltonian
structures and the corresponding Miura transformations of KP hierarchy,
on one hand, and the discrete Toda-like integrable models living on
{\em refinements} of the original lattice connected with the underlying
multi-matrix models, on the other hand.
We show that the connection with the lattice integrable models can be used
to completely characterize the higher KP Hamiltonian structures in terms of
the Darboux-Poisson canonical pairs of free fields.
As it is here explained, there exists an explicit and simple link
between a set of sub-lattice (refined-lattice) spectral equations and the
Lax operators of the multi-boson KP hierarchies expressed
in terms of Poisson-abelian fields with respect to the given Hamiltonian
structure.
The fact, that the KP Hamiltonian structures turn out to be correlated with
the lattice spacing of the corresponding discrete integrable systems,
points to the relevance of the notion of the
discrete lattice formulation for the discussion of the origin of the
multi-hamiltonian structures in continuum integrable models.

Among other results we are obtaining here, is a generalization of the
two-boson realization of the nonlinear, ({\sl i.e.}, non-Lie) $\hWinf$ algebra
\ct{yu-wu}
to a series of $\hWinf$ representations in terms of arbitrary even number of
ordinary free bosonic fields\foot{These representations of $\hWinf$ are
{\em not} equivalent to the representations proposed in \ct{yu} which were
constructed in terms of odd number of scalar fields with alternating
signatures.}. The latter algebra plays an important r{\^{o}}le
as a ``hidden'' symmetry algebra in string-theory-inspired
models with black hole solutions \ct{yu-wu}.
\lskip
{\large {\bf 2. Some Known Basic Results on Reduced KP Hierarchies}}
\lskip
To set the scene we start with a brief recapitulation of some basic
properties of the two-boson KP hierarchy, as well as of multi-boson KP
hierarchies w.r.t. the first Hamiltonian structure,
emphasizing features relevant for the present work.
\lskip
{\bf 2.1 Two-boson KP Hierarchy}
\lskip
We first consider truncated elements of KP hierarchy
of the type $L_{ab} =D + a \(D - b \)^{-1}$, where $D= \pa/ \pa x$
and where we have introduced two Bose currents $(a,b )$ \ct{2boson}.
The Lax operator can be cast in the standard form
$L_{ab} =D + \sumi{n=0} w_n D^{-1-n}$
with coefficients $w_n = (-1)^n a (D - b)^n \cdot 1$
written in terms of the \faa polynomials.
A calculation of the Poisson bracket structures using definitions\foot{Here
and below the following notations are used. ${\Tr}_A Z \equiv
\int dx \, {\rm Res} Z = \int dx \, Z_{-1}(x)\,$ is the Adler trace for
arbitrary pseudo-differential operator $Z = \sum_{k \geq -\infty} Z_k (x) D^k
\,$, and the subscripts $\pm$ in $Z_{\pm}\,$ denote taking the purely
differential or the purely pseudo-differential part of $Z$, respectively.}:
\br
{\pbbr{\me{L}{X}}{\me{L}{Y}}}_1 \eq
- \llangle L \bv \left\lb X,\, Y \right\rb \rrangle  \lab{first-KP}\\
{\pbbr{\me{L}{X}}{\me{L}{Y}}}_2 \eq {\Tr}_A \( \( LX\)_{+} LY -
\( XL\)_{+} YL \) \nonu  \\
&+& \!\!\int dx \, {\rm Res}\Bigl( \sbr{L}{X}\Bigr) \pa^{-1}
{\rm Res}\Bigl( \sbr{L}{Y}\Bigr) \qquad     \lab{second-KP}
\er
yields the first bracket structure  of two-boson $(a,b)$ system to be given
by $\pbr{a(x)}{b(y)}_1= - \d^{\pr} (x-y)$ and zero otherwise.
This leads the coefficients $w_n$ of $L_{ab}$ to satisfy Poisson-bracket
structure of the linear $\Win1\,$ algebra type.
The second bracket structure \rf{second-KP} takes in this case the form:
\br
\{ a (x) \, , \, b (y) \}_2 &=& -b(x) \d^{\pr} (x-y) - \d^{\pr\pr} (x-y)
\nonu\\
\{ a (x) \, , \, a (y) \}_2 &= &  -2 a (x) \d^{\pr} (x-y) -a^{\pr}
(x) \d (x-y) \lab{2pab}\\
\{ b (x) \, , \, b (y) \}_2 &=&- 2\, \d^{\pr} (x-y) \nonu
\er
and based on this bracket $w_n$ satisfy the $\hWinf\,$ algebra.

The above two-boson hierarchy is gauge-equivalent
to the model based on the pseudo-differential operator \ct{YW9111}:
\be
L_{ce} = \( D- e \)\( D- c\)\( D- e-c \)^{-1}
= D + \( e^{\pr} + ec \) \( D- e-c \)^{-1}
\lab{celax}
\ee
The Miura-like connection between these hierarchies
generalizes the usual Miura transformation
between one-bose KdV and mKdV structures and takes a form \ct{AFGMZ}:
\be
a = e^{\pr} + ec \qquad ; \qquad b= e+c
\lab{gmiura}
\ee
This Miura transformation can easily be seen to abelianize the second bracket
\rf{2pab}, meaning that:
\be
\pbr{e\,(x)}{c\,(y)}_2 \, =\,- \d^{\pr} (x-y)
\lab{p2ec}
\ee
and $a,b$ as given by \rf{gmiura} satisfy \rf{2pab} by construction.

The above structures naturally appear in connection with the Toda and Volterra
lattice hierarchies \ct{Toda}. Consider namely the spectral equation
(here $\pa \equiv \pa/\pa t_{1,1}$ where $t_{1,1}$ denotes the first evolution
parameter) :
\br
\pa \Psi_n \eq \Psi_{n+1} + a_0 (n) \Psi_n     \lab{psia}\\
\l \Psi_n \eq \Psi_{n+1}  + a_0 (n) \Psi_n
+ a_1 (n) \Psi_{n-1}  \lab{psic}
\er
We can cast \rf{psic} in the form $\l \Psi_n = L_n^{(1)} \Psi_{n}$ with
$L_n^{(1)} \equiv \pa + a_1 (n) \(\pa - a_0 (n-1)\)^{-1} $.
Connection to the continuous hierarchy is now established by setting
$a_1 (n) = a$ and $a_0 (n-1) = b$.
The Miura transformed hierarchy described by \rf{celax} can be associated
with ``square-root'' lattice of the original Toda lattice system of \rf{psic}:
\be
\l^{1/2}\; {\wti \Psi}_{n+\h} = \Psi_{n+1} + \cA_{n+1} \Psi_n
\quad;\quad
\l^{1/2} \; \Psi_n = {\wti \Psi}_{n+\h} + \cB_n {\wti \Psi}_{n-\h}
\lab{sqra}
\ee
which defines the Volterra chain equations \ct{Toda}.
Also \rf{sqra} can be cast into $\l \Psi_n = L_n^{(1)} \Psi_{n}$ form with
\be
L_n^{(1)} =
\( \pa - \cA_n \) \( \pa - \cB_{n-1} \) \( \pa - \cB_{n-1} - \cA_{n} \)^{-1}
\lab{ncelax}
\ee
which upon identification $\cA_n = e, \cB_{n-1}= c$ agrees with \rf{celax}.
Furthermore using one of the Volterra equations $\pa \cA_n =
\cA_n (\cB_n - \cB_{n-1} ) $ we can rewrite \rf{ncelax} as
\be
L^{(1)}_n = \( \pa - \cA_n \) \( \pa - \cB_{n} - \cA_{n} \)^{-1}
\( \pa - \cB_{n} \) = \pa + \cB_n \( \pa - \cB_{n} - \cA_{n} \)^{-1} \cA_n
\lab{wy}
\ee
which upon identification $\cB_n  = \bsj\, ,\,
\cA_n = \sj $ takes the form $L= D + \bsj\, \(D - \sj \,- \bsj \,\)^{-1} \sj$
in which the so-called quadratic two-boson KP hierarchy appeared in connection
with $SL(2,\IR)/U(1)$ coset model \ct{YW9111}.

The above two simple 2-boson models will be generalized in the next two
sections to the arbitrary multi-boson KP hierarchies.
\lskip
{\bf 2.2 Multi-Boson KP Hierarchy: the First Bracket.}
\lskip
Let us now write the generalization of the two-boson Lax $L_{ab}$ to the
arbitrary $2M$-field Lax operator \ct{BX9212} :
\be
L_{M} = D + \sum_{l=1}^{M} \AM{l}
\( D - \BM{l}\)^{-1} \( D - \BM{l+1}\)^{-1} \cdot\cdot\cdot
\( D - \BM{M}\)^{-1}    \lab{3-3}
\ee

The main result of the investigation presented in \ct{ANP93}
is contained in the following:
\lskip
\prop {\em The $2M$-field Lax operators
\rf{3-3} are consistent Poisson reductions
of the full KP Lax operator
for any $\, M=1,2,3,\ldots \,$} .

The proof was based on the recursive
formula valid for arbitrary $M=2,3,\ldots$
\br
L_{M} &\equiv& L_M (a,b) \equiv L_M \(a_1 ,b_1 ; \ldots
; a_M ,b_M \)  \nonu  \\
L_{M} \eq e^{\int b_{M}} \Bigl\lb b_{M} +
(a_{M} - a_{M-1} )D^{-1} + D L_{M-1} D^{-1} \Bigr\rb
e^{-\int b_{M}}                   \lab{3-1}
\er
which describes the $2M$-field Lax operators in terms of
the boson fields $\, \( a_r , b_r \)_{r=1}^M \,$ spanning
Heisenberg Poisson bracket algebra :
\be
\lcurl a_r (x),\, b_s (y) \rcurl_{P^{\pr}} =
 - \d_{rs} \pa_x \d (x-y)          \lab{3-2}
\ee

In other words, we proved that the  first Poisson
bracket structure for $L_{M}$ from \rf{3-3} is given by:
\be
\Bigl\{ \llangle L_{M} \bv X \rrangle \, ,\,
\llangle L_{M} \bv Y \rrangle \Bigr\}_{P^{\pr}} =
- \llangle L_{M} \bv \left\lb X,\, Y
\right\rb \rrangle    \lab{3-6}
\ee
where $\, X,\, Y\,$ are arbitrary fixed elements of the algebra
of pseudo-differential operators and $\, \langle \cdot \v \cdot
\rangle = {\Tr}_A (\cdot \,\cdot )\,$ indicates the Adler bilinear pairing.
The subscript $P^{\pr}$ in \rf{3-6} indicates that the constituents of
$L_M (a,b)$ satisfy \rf{3-2}.

As a result of \rf{3-1} the coefficient fields of the Lax operator \rf{3-3}
satisfy themselves the recursion relations :
\br
\AM{M} \eq a_M \quad, \quad \BM{M} = b_M \quad ,\quad
\BM{l} = b_M + B^{(M-1)}_l \quad \; (l=1,2,\ldots ,M-1)
\lab{recur}  \\
A_{1}^{(M)}\eq \( \pa + B_{1}^{(M-1)}\)  A_{1}^{(M-1)} \; ,\quad
\AM{l} = A_{l-1}^{(M-1)} + \( \pa + B_l^{(M-1)} \) A_l^{(M-1)}
\quad  (l=2,\ldots ,M-1)  \nonu
\er
These recursion relations can be easily solved in terms of the
free fields $\, \( a_r , b_r \)_{r=1}^M \,$ from \rf{3-2} to yield :
\br
\BM{l} \eq \sum_{s=l}^{M} b_s \qquad \qquad , \qquad \qquad
\AM{M} = a_{M}
\lab{3-4}  \\
\AM{M-r} \eq \sum_{n_r =r}^{M-1} \cdot\cdot\cdot \sum_{n_2 =2}^{n_3 -1}
\sum_{n_1 =1}^{n_2 -1} \( \pa + b_{n_r} + \cdot\cdot\cdot +
b_{n_r -r +1} \) \cdot\cdot\cdot \( \pa + b_{n_2} + b_{n_2 -1}\)
\( \pa + b_{n_1}\) a_{n_1} \lab{3-5}
\er
\lskip
{\large {\bf 3. Multi-Boson KP Hierarchy: the Second Bracket.}}
\lskip
{\bf 3.1 Generalized Miura Transformation for Multi-Boson KP Hierarchies}
\lskip
The generalized Miura transformation for multi-boson KP hierarchies, which we
are going to construct in this section, can be
viewed as ``abelianization'' of the {\em second} KP Hamiltonian structure
\rf{second-KP},
{\sl i.e.}, expressing the coefficient fields of the pertinent KP Lax operator
in terms of canonical pairs of ``Darboux'' fields. The explicit construction
relies on the following recurrence relation for multi-boson KP Lax operators :
\br
L_{M} \equiv L_M (c,e) \equiv L_M \(c_1 ,e_1 ; \ldots
; c_M ,e_M \) \qquad \qquad  \nonu  \\
L_{M} = e^{\int c_{M}} \( D + c_{M} - e_{M} \) L_{M-1}
\( D - e_{M} \)^{-1} e^{-\int c_{M}} \lab{1a} \\
M=1,2 \ldots \quad , \quad L_0 \equiv D \phanta \nonu \\
\lcurl c_k(x) \, ,\, e_l(y) \rcurl = - \d_{kl} \pa_x \d (x-y) \quad , \;
k,l =1,2, \ldots, M \lab{1b}
\er
As it will be shown below, the pairs
$(c_r ,e_r )_{r=1}^{M}\,$ are the ``Darboux'' canonical pairs for
the second KP bracket for arbitrary $M$.
This defines a sequence of the multi-boson KP Lax operators in terms of the
Darboux-Poisson pairs with respect to the second bracket, very much like
\rf{3-1} defined a similar sequence of Lax operators in terms of the
Darboux-Poisson pairs with respect to the first bracket.

Eq.\rf{1a} implies the following recurrence relations for the coefficient
fields of \rf{3-3} :
\br
B^{(M)}_k = B^{(M-1)}_k + c_{M} \quad 1 \leq k \leq M-1 \quad ,
B^{(M)}_{M} = c_{M} + e_{M}  \lab{2a}  \\
A^{(M)}_1 = \( \pa + B^{(M-1)}_1 + c_{M} - e_{M} \) A^{(M-1)}_1 \lab{3a} \\
A^{(M)}_k = A^{(M-1)}_{k-1} +
\( \pa + B^{(M-1)}_k + c_{M} - e_{M} \) A^{(M-1)}_k
\quad ,\quad 2 \leq k \leq M-1   \lab{4a} \\
A^{(M)}_{M} = A^{(M-1)}_{M-1} + \( \pa + c_{M}\) e_{M} \qquad \quad \lab{5a}
\er
\lskip
(1) Example -- 2-boson KP :
\br
L_1 = e^{\int c_1} \( D + c_1 - e_1 \) D \( D - e_1 \)^{-1} e^{-\int c_1} =
D + A^{(1)}_1 \( D - B^{(1)}_1 \)^{-1} \lab{7} \\
A^{(1)}_1 = \( \pa + c_1\) e_1 \quad ,\quad  B^{(1)}_1 = c_1 + e_1  \lab{8}
\er
Here we recognize the structure of the two-boson hierarchy from \rf{celax} as
well the generalized Miura map \rf{gmiura}.
\lskip
(2) Example -- 4-boson KP :
\br
L_2 = e^{\int c_2} \( D + c_2 - e_2 \)
\lb D + A^{(1)}_1 \( D - B^{(1)}_1 \)^{-1} \rb
\( D - e_2 \)^{-1} e^{-\int c_2}  \nonu  \\
= D + A^{(2)}_2 \( D - B^{(2)}_2 \)^{-1}
+ A^{(2)}_1 \( D - B^{(2)}_1 \)^{-1} \( D - B^{(2)}_2 \)^{-1}  \lab{9} \\
A^{(2)}_2 = A^{(1)}_1 + \( \pa + c_2\) e_2  =
\( \pa + c_1\) e_1 + \( \pa + c_2 \) e_2  \quad   \lab{10a} \\
A^{(2)}_1 = \( \pa + B^{(1)}_1 + c_2 - e_2 \) A^{(1)}_1
= \( \pa + e_1 + c_1 + c_2 -e_2 \) \( \pa + c_1 \) e_1  \lab{11a} \\
B^{(2)}_2 = c_2 + e_2  \quad , \quad B^{(2)}_1 = B^{(1)}_1 + c_2
= e_1 + c_1 + c_2  \qquad \lab{10b}
\er
where $A^{(1)}_1$ and $B^{(1)}_1$ are substituted with their expressions
{}from \rf{8}.
It is easy to derive a second bracket structure for the above fields
directly from \rf{1b}. A simple calculation gives
$\pbr{B^{(2)}_2 (x)}{B^{(2)}_2 (y)}
= \pbr{B^{(2)}_1 (x)}{B^{(2)}_1 (y)} = - 2 \pa_x \d (x-y)$,
$\pbr{B^{(2)}_2 (x)}{B^{(2)}_1 (y)}= -  \pa_x \d (x-y)$ etc. thus reproducing
the result of \ct{BX9305} based on Lenard relations.

{}From recursive relation \rf{1a} we can obtain
closed expressions for the arbitrary Lax $L_{M}$ for $M=1,2,\ldots$
directly in terms of the building blocks $(c_r ,e_r )_{r=1}^{M}$ :
\be
L_{M} = \( D - e_{M} \) \prod_{k=M-1}^1 \( D - e_k - \sum_{l=k+1}^{M} c_l \)
\( D - \sum_{l=1}^{M} c_l \) \prod_{k=1}^{M}
\( D - e_k - \sum_{l=k}^{M} c_l \)^{-1}  \lab{1c}
\ee
or, equivalently, in a ``dressing'' form:
\br
L_{M} = \cU_{M} \cU_{M-1} \ldots \cU_1 \, D \, \cV_1^{-1} \ldots
\cV_{M-1}^{-1} \cV_{M}^{-1}   \lab{1d} \\
\cU_k \equiv \( D - e_k \) e^{\int c_k}  \qquad , \qquad
\cV_k \equiv e^{\int c_k} \( D - e_k \) \quad  \lab{1e}
\er

The recurrence relations \rf{2a}--\rf{5a} can be explicitly solved in terms of
the Darboux fields :
\br
B^{(M)}_k \eq e_k +\sum_{l = k}^{M}  c_l \quad , \quad
1 \leq k \leq M \qquad A^{(M)}_{M} =\sum_{k=1}^{M}
\( \pa +c_k \) e_k \lab{11}  \\
A^{(M)}_k \eq \sum_{n_k =1}^k \( \pa + e_{n_k} - e_{n_k +M-k}
+ \sum_{l_k = n_k}^{n_k +M-k} c_{l_k} \) \times \nonu  \\
&\times& \!\!  \sum_{n_{k-1}=1}^{n_k} \( \pa + e_{n_{k-1}} - e_{n_{k-1} +M-1-k}
+ \sum_{l_{k-1} = n_{k-1}}^{n_{k-1} +M-1-k} c_{l_{k-1}} \)
\times \cdots  \nonu  \\
&\times& \!\! \sum_{n_2 =1}^{n_3} \( \pa + e_{n_2} - e_{n_2 +1} +
c_{n_2} + c_{n_2 +1} \) \, \sum_{n_1 =1}^{n_2} \( \pa + c_{n_1} \) e_{n_1}
\quad , \;\; k=1, \ldots, M-1 \lab{14}
\er
The Miura-transformed form of $L_{M}$ reads explicitly :
\br
&&\!\!L_{M} = D + \sum_{k=1}^{\infty}
U_k \lb (c,e) \rb (x) D^{-k}   \lab{rec-A}  \\
&&\!\!U_k \lb (c,e) \rb (x) = P^{(1)}_{k-1} \( e_{M} + c_{M} \)
\sum_{l=1}^{M} \( \pa + c_l \) e_l \;\, +     \lab{rec-B} \\
&&\!\!\sum_{r=1}^{\min (M-1,k-1)} A_{M-r}(c,e) P^{(r+1)}_{k-1-r}
\bigl( e_{M} + c_{M} , e_{M-1} + c_{M-1} + c_{M} , \ldots ,
e_{M-r} + \sum_{l=M-r}^{M} c_l \bigr)  \nonu
\er
where $\,A_{M-r}(c,e)\,$ are the same as in \rf{14}, and
$\, P^{(N)}_n \,$ denote the (multiple) \faa polynomials \ct{ANP93} :
\be
P^{(N)}_n (B_N ,B_{N-1},\ldots ,B_1) =
\sum_{m_1 + \cdots + m_N = n}
\( -\pa + B_1 \)^{m_1} \cdots \( -\pa + B_N\)^{m_N}
\cdot 1   \lab{multifaa}
\ee

Now, upon substitution of \rf{rec-A}--\rf{rec-B} into \rf{second-KP},
we obtain a series of explicit (Poisson bracket) realizations of the nonlinear
$\hWinf$ algebra in terms of $2M$ bosonic fields,
satisfying \rf{1b}, for any $M=1,2, \ldots$ .
\lskip
{\bf 3.2 Consistency of Multi-Boson KP Poisson Reduction w.r.t. the
Second Bracket}
\lskip
The main result of this paper follows from the following general
statement (analogue of the Kupershmidt-Wilson theorem for mKdV \ct{Dickey}).
\lskip
\prop {\em Let $(c_r ,e_r )_{r=1}^{M}\,$ obey the Heisenberg Poisson
algebra \rf{1b}. Then the reduced Lax operators $L_{M}\,$ \rf{1d} (or
\rf{rec-A}--\rf{rec-B}) satisfy the second KP Hamiltonian structure
\rf{second-KP}.}

\proof It proceeds by induction w.r.t. $M$. The case $M=1$ is the familiar
example of two-boson KP hierarchy. Substituting according to
\rf{1d}--\rf{1e} $\, L_{M}= \cU_{M} L_{M-1} \cV^{-1}_{M}$ into \rf{second-KP}
we have :
\br
\pbbr{\me{L_{M}}{X}}{\me{L_{M}}{Y}} =  \phantb  \nonu \\
\pbbr{\me{L_{M-1}}{\cV^{-1}_{M}X\cU_{M}}}
{\me{L_{M-1}}{\cV^{-1}_{M}Y\cU_{M}}} +  \phantb \lab{pb-A} \\
\pbbr{\me{\cU_{M}}{L_{M-1} \cV^{-1}_{M}X}}{\me{\cU_{M}}{L_{M-1}\cV^{-1}_{M}Y}}
+ \pbbr{\me{\cV^{-1}_{M}}{X\cU_{M}L_{M-1}}}{\me{\cV^{-1}_{M}}{Y\cU_{M}L_{M-1}}}
+  \lab{pb-B} \\
\pbbr{\me{\cU_{M}}{L_{M-1} \cV^{-1}_{M}X}}{\me{\cV^{-1}_{M}}{Y\cU_{M}L_{M-1}}}
+ \pbbr{\me{\cV^{-1}_{M}}{X\cU_{M}L_{M-1}}}{\me{\cU_{M}}{L_{M-1}\cV^{-1}_{M}Y}}
\lab{pb-C}
\er
In all terms \rf{pb-A}--\rf{pb-C} in the r.h.s. of the above equation
it is understood that the Poisson brackets are taken
w.r.t. the left members in the angle brackets.
Henceforth, for simplicity, we shall skip the subscripts $M$ of $\,\cU ,\cV ,
c, e\,$. Using the induction hypothesis
and simple identities for pseudo-differential operators, the bracket \rf{pb-A}
takes the form :
\br
{\rm \rf{pb-A}} = {\Tr}_A \( \( L_{M-1} \cV^{-1} X \cU \)_{+} L_{M-1} \cV^{-1}
Y \cU -
\( \cV^{-1} X \cU L_{M-1}\)_{+} \cV^{-1} Y \cU L_{M-1} \) + \qquad \nonu  \\
\int dx \, {\rm Res}\Bigl( \sbr{L_{M-1}}{\cV^{-1}X\cU}\Bigr) \pa^{-1}
{\rm Res}\Bigl( \sbr{L_{M-1}}{\cV^{-1}Y\cU}\Bigr) \qquad \qquad \qquad    \\
= {\Tr}_A \( \( L_{M} X\)_{+} L_{M} Y -
\( X L_{M}\)_{+} Y L_{M} \) -  \phanta \qquad \lab{pb-AAA}  \\
{\Tr}_A \llb \( \cU^{-1} \( L_{M}X\)_{+} \cU \)_{-}
\( \cU^{-1} \( L_{M}Y\)_{+} \cU \)_{+} -
\( \cV^{-1} \( X L_{M}\)_{+} \cV \)_{-}
\( \cV^{-1} \( Y L_{M}\)_{+} \cV \)_{+} \rrb  \lab{pb-A2} \\
+ \int dx \, {\rm Res}\Bigl( \cU^{-1} L_{M}X \cU
-\cV^{-1} X L_{M}\cV \Bigr) \pa^{-1} \Bigl( \cU^{-1} L_{M}Y \cU -
\cV^{-1} Y L_{M}\cV \Bigr)  \qquad  \lab{pb-A1}
\er
The following identities, valid for arbitrary pseudo-differential
operator $Z$ and arbitrary function $f$, will be used in the sequel :
\br
{\rm Res} \( Z - \cU^{-1} Z\cU \) = \pa_x {\rm Res} \( \( D -e\)^{-1} Z \)
\quad ,\quad
{\rm Res} \( \( D-f\)^{-1} Z \) = \( e^{\int f} Z e^{- \int f} \)_0^{(R)}
\nonu  \\
{\rm Res} \( Z - \cV^{-1} Z\cV \) =
\pa_x {\rm Res} \( \( D -e-c\)^{-1} Z \)   \phantb \lab{Res-ident}
\er
where the last sub/superscripts indicate taking the zero-order (c-number)
part of the corresponding {\em right-ordered}\foot{That means, the
 coefficient functions are to the right w.r.t. differential operators $D$.}
pseudo-differential operator.
Using \rf{Res-ident} the term \rf{pb-A1} can be rewritten in the form :
\br
{\rm \rf{pb-A1}} = \int dx \, {\rm Res}\Bigl( \sbr{L_{M}}{X}\Bigr) \pa^{-1}
{\rm Res}\Bigl( \sbr{L_{M}}{Y}\Bigr) +  \qquad \qquad \nonu  \\
\int dx \, {\rm Res}\Bigl( \sbr{L_{M}}{X}\Bigr)
{\rm Res}\( \( D-e-c\)^{-1} Y L_{M} - \( D-E\)^{-1} L_{M} Y \) -
\Bigl( \, X \longleftrightarrow Y \, \Bigr)   \nonu \\
+ \int dx \; \pa_x \lcurl {\rm Res}\( \( D-e-c\)^{-1} X L_{M} -
\( D-e\)^{-1} L_{M} X \) \rcurl  \qquad \nonu \\
\times {\rm Res}\( \( D-e-c\)^{-1} Y L_{M} - \( D-e\)^{-1} L_{M} Y \)
\qquad \lab{pb-A1a}
\er
Similarly, taking into account \rf{Res-ident}, the trace term \rf{pb-A2} takes
the form :
\br
{\rm \rf{pb-A2}} = \int dx \, \pa_x \lcurl {\rm Res} \( \( D-e\)^{-1}
L_{M}X \) \rcurl \, {\rm Res} \( \( D-e\)^{-1} L_{M}Y \) \qquad \nonu \\
- \int dx \, \pa_x \lcurl {\rm Res} \( \( D-e-c\)^{-1} XL_{M}\) \rcurl \,
{\rm Res} \( \( D-e-c\)^{-1} YL_{M} \) \qquad \lab{pb-AAB}
\er
Finally, using the Heisenberg Poisson algebra \rf{1b} for
$\( c_{M},e_{M} \)$, (part of assumption in the Proposition), and using
again \rf{Res-ident}, the sum of the terms \rf{pb-B} and \rf{pb-C} reads :
\br
{\rm \rf{pb-B}} + {\rm \rf{pb-C}} =
\int dx \, \llb \pa_x \lcurl{\rm Res}\( \( D-e\)^{-1} L_{M}X \)\rcurl  -
{\rm Res}\Bigl( \sbr{L_{M}}{X}\Bigr) \rrb \times \nonu \\
\times {\rm Res}\( \( D-e-c\)^{-1} Y L_{M} - \( D-e\)^{-1} L_{M} Y \)
- \Bigl( \, X \longleftrightarrow Y \, \Bigr) \lab{pb-BC}
\er

Thus, collecting the results \rf{pb-AAA}, \rf{pb-A1a}, \rf{pb-BC}
and \rf{pb-AAB}, we achieve precise cancellation of the unwanted terms
leaving us with the desired result :
\br
\pbbr{\me{L_{M}}{X}}{\me{L_{M}}{Y}} \eq
{\Tr}_A \( \( L_{M}X\)_{+} L_{M}Y - \( XL_{M}\)_{+} YL_{M} \)
\nonu  \\
&+& \!\! \int dx \, {\rm Res}\Bigl( \sbr{L_{M}}{X}\Bigr) \pa^{-1}
{\rm Res}\Bigl( \sbr{L_{M}}{Y}\Bigr) \nonu
\er
\lskip
{\large {\bf 4. The Sub-Lattices and Abelianization of Higher Brackets}}
\lskip

In this section the recurrence relations presented above will be
traced back to the lattice
formulation based on the discrete spectral equations.
We start by providing link between lattice formulation
and recurrence relation in \ct{ANP93}
amounting to abelianization of the first KP bracket.

We start with the spectral equation:
\be
\l \Psi_n =  L_n^{(N)} \Psi_{n} \qquad \forall n
\lab{spectr}
\ee
where
\be
L_n^{(N)} = \pa + \sum_{k=1}^N a_k (n) {1 \ov \pa - a_0 (n-k)}
\ldots {1 \ov \pa - a_0 (n-1)}
\lab{lnn}
\ee
Multiplying $L_n^{(N)}$ on both sides by $1= e^{\int a_0 (n-1)}
e^{-\int a_0 (n-1)}$ we arrive at:
\br
L_n^{(N)} \eq e^{\int a_0 (n-1)} \lcurl a_0 (n-1) + a_1 (n) \pa^{-1}
+\( \pa   \right.\right.\lab{ida}\\
\!\!&+&\!\!  \left.\left.\sum_{k=2}^N a_k (n) {1 \ov \pa +a_0 (n-1)-
a_0 (n-k)} \ldots {1 \ov \pa +a_0 (n-1)- a_0 (n-2)} \pa^{-1} \) \rcurl
e^{-\int a_0 (n-1)}  \nonu
\er
Recalling the Toda equation of motion, which is a consistency condition
following from \rf{psia} and \rf{spectr}:
\be
a_k (n) = a_{k} (n-1) + \biggl(\pa + a_0 (n-k) - a_0 (n-1) \biggr)
a_{k-1} (n-1) \quad \; k=1,2,\ldots,n
\lab{toda}
\ee
and the simple identity:
\br
\lb\pa a_{k-1} (n-1) \!\!&-&\!\! a_{k-1} (n-1) \( a_0 (n-1) - a_0 (n-k) \)
\rb {1 \ov \pa +a_0 (n-1)- a_0 (n-k)}\nonu \\
&=&\!\!\!\! \pa \( a_{k-1} (n-1) {1 \ov \pa +a_0 (n-1)- a_0 (n-k)} \)
-a_{k-1} (n-1) \lab{idb}
\er
we find
\br
L_n^{(N)} \eq e^{\int a_0 (n-1)} \lcurl a_0 (n-1) + a_1 (n) \pa^{-1} \right.
\lab{idc} \\
&+&\!\! \lb a_{N} (n-1) {1 \ov \pa +a_0 (n-1)- a_0 (n-N)} \ldots
{1 \ov \pa +a_0 (n-1)- a_0 (n-2)} - a_{1} (n-1) \rb \pa^{-1}
\nonu\\
&+&\!\! \left.
\pa \lb\pa + \sum_{l=1}^{N-1} a_l (n-1) {1 \ov \pa +a_0 (n-1)- a_0 (n-1-l)}
\ldots {1 \ov \pa +a_0 (n-1)- a_0 (n-2)}  \rb \pa^{-1}\rcurl\nonu \\
&\times&\!\!  e^{-\int a_0 (n-1)} \nonu
\er
This connects lattice formulation to recurrence relation \rf{3-1} \ct{ANP93}.
To see it more clearly recall now expression \rf{3-3} for the continuous Lax
operator. Comparing with \rf{lnn} we see the following correspondence:
\be
A^{(N)}_{N-k+1} \sim a_k (N) \quad;\quad A^{(N)}_{N} = a_N \sim a_1(N)
\quad;\quad B^{(N)}_{N-k+1} \sim a_0 (N-k)
\quad\;k = 1,\ldots,N
\lab{corres}
\ee
where we put $n=N$.
It is now obvious that the lattice Toda eq. of motion \rf{toda}
corresponds to the recurrence relation \rf{recur}:
\be
A^{(N)}_{l} = A_{l-1}^{(N-1)} + \( \pa + B_l^{(N-1)} \) A_l^{(N-1)}
\quad  (l=2,\ldots ,N-1)  \lab{rec1}
\ee
with $
B^{(N)}_{l} = b_M + B^{(N-1)}_l$, $B^{(N)}_{N} = b_N \sim a_0 (N-1)$
and the further correspondence $B^{(N-1)}_l \sim  a_0 (l-1) -a_0 (n-1)$
established by comparing \rf{3-3} with \rf{idc}.

Therefore we have established relation between the lattice system with the
Toda eq. \rf{toda} relating functions on different sites
and corresponding continuous system with recurrence relations relating
different orders of reduction of KP.

We now discuss a link between ``square-root'' lattice formulation and
recurrence
relation \rf{1a}.
We define spectral equation:
\br
\l^{1/2}\; {\wti \Psi}_{n+\h} \eq \Psi_{n+1} + \cA^{(0)}_{n+1} \Psi_n +
\sum_{p=1}^N \cA^{(p)}_{n-p+1} \Psi_{n -p} \lab{sqraa} \\
\l^{1/2} \; \Psi_n \eq {\wti \Psi}_{n+\h} + \cB^{(0)}_n {\wti \Psi}_{n-\h}
\lab{sqrbb}
\er
We also have time evolution eqs:
\be
{\wti \Psi}_{n+\h} = \( \pa - \cB^{(0)}_n - \cA^{(0)}_{n} \){\wti \Psi}_{n-\h}
\quad;\quad
\Psi_{n+1} = \( \pa - \cB^{(0)}_n - \cA^{(0)}_{n+1} \) \Psi_{n}
\lab{evolution}
\ee
We therefore find
\be
\l^{1/2}\; {\wti \Psi}_{n+\h}= \( \pa - \cB^{(0)}_n + \sum_{p=1}^N
\cA^{(p)}_{n-p+1} ( \pa - \cB^{(0)}_{n-p} - \cA^{(0)}_{n-p+1} )^{-1}
\cdots ( \pa - \cB^{(0)}_{n-1} - \cA^{(0)}_{n} )^{-1} \) \Psi_n
\lab{speca}
\ee
and
\be
\l^{1/2}\; \Psi_{n}= \( \pa - \cA^{(0)}_n \) {\wti \Psi}_{n-\h}
\lab{specb}
\ee
{}from the last two relations we find:
\br
\l\; \Psi_{n}\eq\( \pa - \cA^{(0)}_n \)
\( \pa - \cB^{(0)}_{n-1} + \sum_{p=1}^N
\cA^{(p)}_{n-p} ( \pa - \cB^{(0)}_{n-p-1} - \cA^{(0)}_{n-p} )^{-1}
\cdots ( \pa - \cB^{(0)}_{n-2} - \cA^{(0)}_{n-1} )^{-1} \) \nonu \\
&&( \pa - \cB^{(0)}_{n-1}-\cA^{(0)}_{n} )^{-1} \Psi_n
\lab{almost}
\er
This defines a Lax operator through $ \l \Psi_n =L^{(N+1)}_n
\Psi_{n}$ where
\be
L^{(N+1)}_n = e^{\int \cB^{(0)}_{n-1}} \( \pa - \cA^{(0)}_n +\cB^{(0)}_{n-1}\)
L^{(N)}_n \( \pa - \cA^{(0)}_n \)^{-1} e^{-\int \cB^{(0)}_{n-1}}
\lab{recu}
\ee
and
\be
L^{(N)}_n=  \pa  + \sum_{p=1}^N
\cA^{(p)}_{n-p} ( \pa + \cB^{(0)}_{n-1}- \cB^{(0)}_{n-p-1} -
\cA^{(0)}_{n-p} )^{-1}
\cdots ( \pa + \cB^{(0)}_{n-1}-\cB^{(0)}_{n-2} - \cA^{(0)}_{n-1} )^{-1}
\lab{lax}
\ee
establishing one to one correspondence with the recurrence
relation \rf{1a}.

We now introduce notion of ``1/4'' lattice, which is a further sub-lattice
of the ``square-root'' lattice associated to the Volterra hierarchy.
Let us first rewrite \rf{sqra} in a compact notation as
\be
\l^{1/2} \; \Phi_n =  \Phi_{n+1} + V_n  \Phi_{n-1}
\lab{square-lat}
\ee
where we introduced a compact notation intrinsic
for the square-lattice system through $V_{2n} = \cB_n ,V_{2n-1} = \cA_{n}$
and $\Phi_{2n} = \Psi_n,\Phi_{2n-1} = {\wti \Psi}_{n-\h}$.
It turns out that the process of ``refinement'' of the lattice can be
continued. We associate the following spectral system:
\be
\l^{1/4} \; {\wti \Phi}_{n+\h} =  \Phi_{n+1} - W_n  \Phi_{n}
\quad;\quad
\l^{1/4} \; \Phi_n =  {\wti \Phi}_{n+\h}+ W_n  {\wti \Phi}_{n-\h}
\lab{quarter-lat}
\ee
to the ``one-quarter'' lattice.
In \rf{quarter-lat} we have introduced the new object $W_n$ which
can be expressed through square-lattice components as
$W_{2n} = \b_n ,W_{2n-1} = \a_{n}$.
Combining two equations of \rf{quarter-lat} one gets again equation
\rf{square-lat} with $V_n = - W_n W_{n-1}$.
Recalling the Volterra equation $\pa V_n = V_n (V_{n+1} - V_{n-1} ) $
we find the evolution equation for $W_n$-field to be
$\pa W_n = W^2_n (W_{n+1} - W_{n-1} ) $ or in components
\be
\pa \a_n = \a^2_n (\b_n - \b_{n-1} )  \qquad;\qquad
\pa \b_n = \b^2_n (\a_{n+1} - \a_{n} )  \lab{quarter-ev}
\ee
Relation $V_n = - W_n W_{n-1}$ translates in components to
\be
\cA_n = -\a_n \b_n + { \a_n^{\pr} \ov \a_n} \qquad;\qquad
\cB_n = - \b_n  \a_{n} \lab{dnls}
\ee
which identifies the new hierarchy, connected with the ``1/4''-lattice,
with the so-called derivative Non-Linear Schr\"odinger (dNLS)
hierarchy \ct{discrete}.
In \ct{discrete} it was in fact shown that for the fields of dNLS
hierarchy $\a_n = -q , \b_n = - r - q^{\pr}/q^2 $ the third Poisson bracket
structure
\be
\{q(x),r(y)\}_3 = \d^{\pr}(x-y) \lab{dnls-bra}
\ee
takes an abelian form.
\lskip
{\large {\bf 5. Outlook}}
\lskip
The theory of integrable lattice systems has a profound geometrical foundation
and found recently new important physical applications. Among other results
obtained in this paper,
we have used the Toda lattice system and the related
discrete integrable systems, living on its sub-lattices,
to ``coordinatize'' the continuum multi-boson KP hierarchies and their
Hamiltonian structures.
It was shown how KP Miura transformations are related to the underlying
lattice structures and how transition to the ``finer'' sub-lattice
provides the right set of abelian field coordinates for the higher
KP Hamiltonian structures.
In view of the significance of higher Hamiltonian structures for the notion of
integrability this emphasizes the need of a general approach to results we have
obtained in this paper.
A completely systematic theory should not only fully explain
the link between the various lattice integrable systems and the higher
Hamiltonian structures, but should also address the relation between the
Toda lattice Poisson brackets (with discrete indices)
and field-theoretical Poisson brackets in continuum  w.r.t.
the first lattice evolution parameter. We plan to address these interesting
issues in the future.

In addition we hope that
the established connection between the higher Hamiltonian
structures and the pertinent Miura transformations of KP hierarchy,
on one hand, and the discrete Toda-like integrable models
on refined lattices, on the other hand,
will also allow to make further progress in understanding the
integrability of the quantum theories.

\lskip
{\bf Acknowledgements.} We are grateful to L.A. Ferreira and A.H. Zimerman
for reading the manuscript and comments.

\end{document}